\documentclass[aps,showpacs,preprintnumbers,amsmath,amssymb]{revtex4}

\usepackage{graphicx}
\usepackage{bm}

\begin{document}

\title{Comment on ``Phase Reduction of Stochastic Limit Cycle Oscillators''}

\author{Hiroya Nakao}
\affiliation{Department of Physics, Kyoto University, Kyoto 606-8502, Japan}

\author{Jun-nosuke Teramae} 
\affiliation{Brain Science Institute, RIKEN, Wako 351-0198, Japan}

\author{G. Bard Ermentrout} 
\affiliation{Department of Mathematics, University of Pittsburgh,
  Pittsburgh, Pennsylvania 15260, USA}

\date{\today}


\maketitle

In a recent Letter, Yoshimura and Arai~\cite{Yoshimura} claimed that
the conventional phase stochastic differential equation (SDE) used
in~\cite{TN,GP,ER} does not give a proper approximation to limit-cycle
oscillators driven by noise, and proposed a modified phase SDE.  Here
we argue that their claim is not always correct; both SDEs are valid
depending on the situation.

Since physical noise has an associated time scale and all oscillators have
a characteristic rate of attraction, which of the two  SDEs is
appropriate depends on the relative sizes of these two scales. 
As a simple example, let us consider the Stuart-Landau (SL) model used
in~\cite{Yoshimura,TN} driven by a colored noise generated by the
Ornstein-Uhlenbeck process (OUP)~\cite{SP}, which is rescaled such that the
amplitude relaxation time explicitly appears while keeping the limit
cycle and its isochrons invariant,
\begin{align}
  \dot{W}(t) = \{ T^{-1} (1 + i c ) + i \omega \} W - T^{-1} (1 + i c)
  |W|^2 W + \sqrt{2\varepsilon} \xi(t),
  \label{SL}
\end{align}
where $W$ is a complex variable representing the oscillator state, $T$
is the relaxation time of the amplitude, $c$ and $\omega$ are
parameters, $\varepsilon$ is the noise intensity, and $\xi(t)$ is 
OUP noise that is applied only to the real component of $W$ for
simplicity.   $\xi(t)$ is Gaussian-distributed,
and its correlation function is given by $\langle \xi(t) \xi(s)
\rangle = \exp\left( - |t-s| / \tau \right) / (2 \tau)$, which
converges to $\delta(t)$ as $\tau \to 0$.  Thus, $\xi(t)$ gives a
colored-noise approximation to the Wiener process~\cite{SP}.

Introducing the amplitude $R = |W|$ and the isochron phase $\phi =
\arg W - c \ln |W|$~\cite{Yoshimura,TN}, Eq.~(\ref{SL}) can be written
as
\begin{align}
  \dot{R}(t) = T^{-1} ( R - R^3 ) + \sqrt{2 \varepsilon} \cos ( \phi + c \ln R) \
  \xi(t),
  \label{R}
\end{align}
\begin{align}
  \dot{\phi}(t) = \omega - \sqrt{2 \varepsilon} R^{-1} \left\{ \sin ( \phi + c \ln
    R) + c \cos ( \phi + c \ln R) \right\} \ \xi(t).
  \label{phi}
\end{align}
It is now clear that $T$ actually determines the relaxation time of
the amplitude $R$.  The limit cycle in the absence of the noise
($\varepsilon=0$) is simply $R(t) \equiv 1$ and $\phi(t) = \omega t +
\mbox{const}$.

Two different SDEs have been previously derived describing this and other
noisy oscillators.  The non-agreement is due to the order in which
the white-noise limit and the phase limit are taken~\cite{Brown}.
The ``conventional'' model obtained by taking
the phase limit in the first has the form:
\begin{equation}
  d\phi(t) = \left[ \omega + \varepsilon Z(\phi) Z'(\phi) \right] dt
  + \sqrt{2 \varepsilon} Z(\phi) dw(t),
  \label{SDE1}
\end{equation}
where the phase sensitivity (or response) function $Z(\phi)= - \sin \phi - c \cos \phi$ in the present example.  Yoshimura and Arai's modified phase model
obtained by taking the white-noise limit first is given by
\begin{equation}
  d\phi(t) = \left[ \omega + \varepsilon \left\{
      Z(\phi) Z'(\phi) + Y(\phi) \right\} \right] dt
  + \sqrt{2 \varepsilon} Z(\phi) dw(t),
  \label{SDE2}
\end{equation}
where the extra term $Y(\phi) = (1 + c^2) \sin(2\phi)/
2$~\cite{Yoshimura}.

To see which of the two reduced phase SDEs~(\ref{SDE1},~\ref{SDE2})
approximates the original noisy SL model Eqs.~(\ref{R},~\ref{phi})
better, we compare the stationary phase probability density functions
(PDFs) obtained by direct Langevin simulations of
Eqs.~(\ref{R},~\ref{phi}) for different pairs of $(\tau, T)$ with the
PDFs obtained from the two phase SDEs~(\ref{SDE1},~\ref{SDE2}) by
numerically solving the corresponding Fokker-Planck equations.  We fix
$\omega = 1$, $c=2$, $\varepsilon = 0.01$, and vary $\tau$ and $T$
keeping $\tau T = 0.001$ constant.

Figure 1(a) shows the stationary phase PDFs obtained for two typical
cases, $T = 0.01 \ll \tau = 0.1$ and $\tau = 0.01 \ll T = 0.1$.  The conventional phase SDE~(\ref{SDE1}) nicely fits the
original model when $T \ll \tau$, whereas the modified phase
SDE~(\ref{SDE2}) is better when $\tau \ll T$.  Figure 1(b) shows
mean-square errors of the approximate PDFs yielded by
SDEs~(\ref{SDE1},~\ref{SDE2}) from the original PDF given by
Eqs.~(\ref{R},~\ref{phi}) as functions of $T (= 0.001 / \tau)$.  It is
clear that the conventional phase SDE~(\ref{SDE1}) gives a better
approximation for $T < T{*}$, while the modified SDE~(\ref{SDE2}) is
better for $T > T{*}$. 

Summarizing, we have demonstrated that the conventional phase
SDE~(\ref{SDE1}) is also a proper approximation to noisy limit cycles
with sufficiently fast amplitude relaxation, which can be used as a
starting point for further analysis.  It is a natural
generalization of the ordinary phase equation driven by smooth signals
(which becomes evident when written in the Stratonovich Langevin
form~\cite{SP}), and it has a practical advantage of being completely
determined by $\omega$ and $Z(\phi)$, which are both experimentally
measurable.  
The results in \cite{Yoshimura} are valid when literally white noise
is given to
limit-cycle oscillators from the outset.  However, white noise is
actually an idealization of physical processes with small but finite time correlations.  One should be careful of competing small time scales
involved in the problem when the white-noise limit is taken.  
For example, near the bifurcation of a limit cycle, the attraction
to the limit cycle is
slow so that the SDE~(\ref{SDE2}) is reasonable. However, far from the
bifurcation or for relaxation oscillators where the time scale $T$ is
very small, the conventional SDE~(\ref{SDE1}) should be used. 

Finally, the above intuitive arguments can be made rigorous by using
a multi-scale or projection-operator
method~\cite{Brown,SP}, which yields a family of effective phase SDEs
depending on the ratio $\tau / T$~\cite{TE}.  It can be shown that the
two SDEs~(\ref{SDE1},~\ref{SDE2}) are actually two extreme cases
corresponding to $T \to 0, \tau >0$ and $T>0, \tau \to 0$,
respectively.

H. N. thanks Yoshiki Kuramoto and Kensuke Arai for useful comments.

\begin{figure}
  \centering
  \includegraphics[width=1.0\hsize,clip]{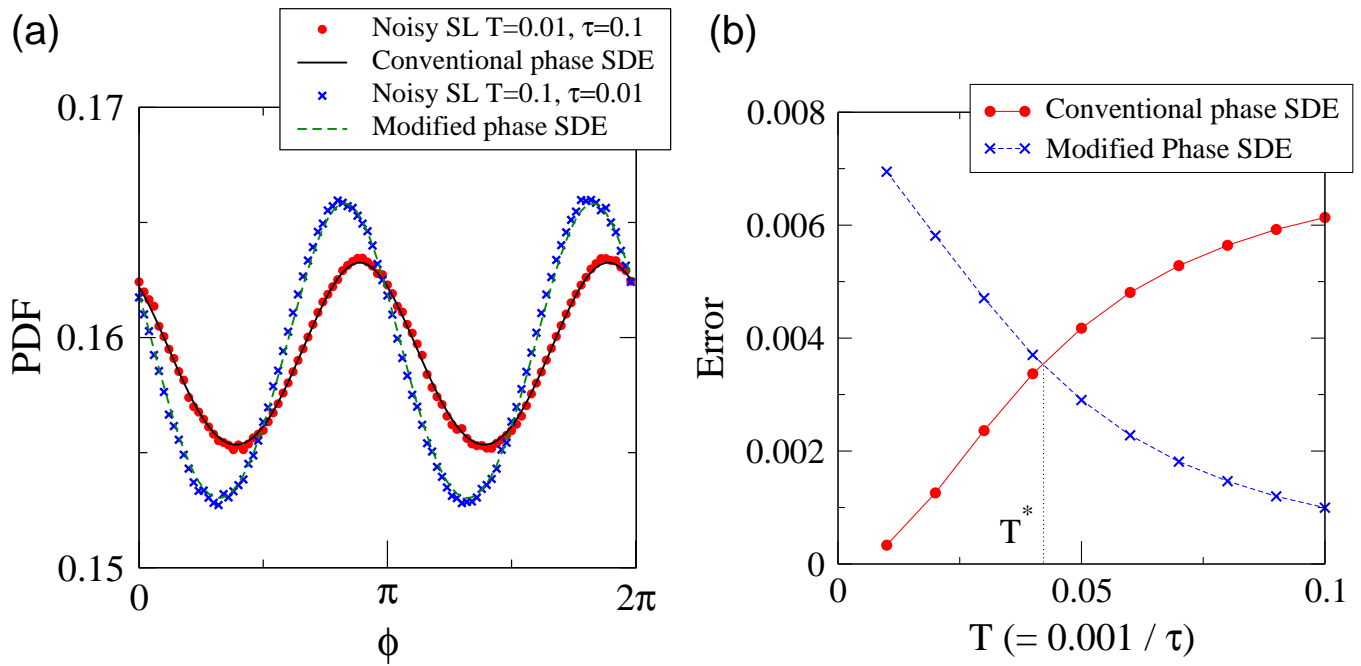}
  \caption{(a) Comparison of stationary phase PDFs obtained directly
    from Eqs.~(\ref{R},~\ref{phi}) with those obtained from the phase
    SDEs~(\ref{SDE1},~\ref{SDE2}) for $(T, \tau) = (0.1, 0.01)$ and
    $(0.01, 0.1)$.  (b) Mean-square errors of the approximate phase
    PDFs of SDEs~(\ref{SDE1},~\ref{SDE2}) from the original PDFs of
    Eqs.~(\ref{R},~\ref{phi}) plotted as functions of the amplitude
    relaxation time $T ( = 0.001 / \tau )$.}
\end{figure}

\end{document}